# Corporate Financial Distress Prediction: Based on Multi-source Data and Feature Selection

Yi Ding [1], Chun Yan [1]

[1] College of Mathematics and Systems Science, Shandong University of Science and Technology, Qingdao, China

**Abstract**

The advent of the era of big data provides new ideas for financial distress prediction. In order to evaluate the financial status of listed companies more accurately, this study establishes a financial distress prediction indicator system based on multi-source data by integrating three data sources: the company's internal management, the external market and online public opinion. This study addresses the redundancy and dimensional explosion problems of multi-source data integration, feature selection of the fused data, and a financial distress prediction model based on maximum relevance and minimum redundancy and support vector machine recursive feature elimination (MRMR-SVM-RFE). To verify the effectiveness of the model, we used back propagation (BP), support vector machine (SVM), and gradient boosted decision tree (GBDT) classification algorithms, and conducted an empirical study on China's listed companies based on different financial distress prediction indicator systems. MRMR-SVM-RFE feature selection can effectively extract information from multi-source fused data. The new feature dataset obtained by selection has higher prediction accuracy than the original data, and the BP classification model is better than linear regression (LR), decision tree (DT), and random forest (RF).

# 1 Introduction

The occurrence of financial distress is predictable (Zeng, 2021). Analyzed at the micro level, the development and financial structure of an enterprise reflect the financial situation of this enterprise, and the traditional financial distress prediction (FDP) research is analyzed for the private information of micro enterprises (Kitowski et al., 2022). The development of big data technology has led to a consequent change in the business environment, and data from multiple sources have laid a strong data foundation for enterprise development (Zhu et al., 2021). Considering the combination of data information from multiple sources may greatly increase the accuracy of forecasting.

However, the rapid development of Internet technology and the complex and changing market environment have made enterprises face more difficult challenges, and there are countless enterprises on the verge of bankruptcy due to financial crises (Liu et al.,2022). In this environment, the study of FDP is crucial to the investment decision of the public, the risk avoidance of enterprises and the supervision of government departments.

Recent studies on FDP have focused on two aspects: the selection of indicators and the selection of models. In terms of indicator selection, the information that can reflect the company's condition is usually used as the predicion indicator, and the early stage mainly focuses on building the model through financial indicators. However, in today's high-speed economic and social development, the accounting information reflected by traditional financial indicators has lagging and manipulability, and distortion of accounting information occurs from time to time (Caserio et al., 2020). Therefore, scholars began to include non-financial indicators to achieve the expansion and improvement of the indicator system (Li et al., 2015; Zhu et al., 2021). Wang et al. (2021) found that the wife's management control is also associated with less debt and more cash, and the wife's empowerment in the firm is associated with lower earnings volatility. Tang et al. (2020) introduced management indicators and text information indicators to construct an indicator system based on traditional financial indicators, and the study showed that the system can improve the prediction accuracy of the model. Some scholars (Guan et al., 2021; Cao et al., 2022) have established targeted indicator systems according to different research objects and research fields, and established FDP indicator systems under different perspectives such as agricultural enterprises, real estate enterprises and low carbon economy. In the context of big data, the factors that enable prediction of corporate financial distress are becoming increasingly diverse. Li et al. (2015) used Extensible Business Reporting Language (XBRL tagging) to achieve judgment on financial status, and Chen et al. (2019) discerned the financial status of listed companies by analyzing the management tone of the textual information disclosed in their annual reports. Some scholars have proposed to use public opinion as an influencing factor, because the emotional tendencies of the public can more truly reflect the company's operating conditions. Song et al. (2015) and Wang (2020) took the shareholders of listed companies as the "sensors" of the enterprise, and tried to establish an FDP model that introduced big data indicators through sentiment analysis processing and financial indicators.

    The research on FDP models has gone through univariate analysis stage, multivariate analysis stage and diversification development stage, multivariate analysis stage and diversification development stage. With the development of artificial intelligence technology, machine learning and other techniques are widely used in the construction of FDP models. Different scholars (Lu et al., 2017; Sankhwar et al., 2020; Sun et al.,2021) have tried and compared the algorithms and models continuously in order to expect a higher correct rate of prediction. Ward et al. (2017) introduced classification tree ensembles (CTEs) to bank financial crisis forecasting and found that

CTEs greatly improved out-of-sample forecasting performance compared to other FDP systems. Song et al. (2021) established a financial risk assessment model based on back propagation neural network (BPNN) and deep belief network (DBN) which greatly improved the accuracy of prediction.

In the selection of FDP indicators, financial crisis data have high-dimensional characteristics. There are many indicators that can reflect the financial situation of the company, but these indicators have a large correlation among them and contain more redundant indicators, which brings many difficulties for the study of FDP (Du et al., 2020). There are two main types of methods for the selection of indicators in existing studies. One is to construct an indicator system by selecting appropriate indicators among alternative indicators based on historical experience and in a qualitative manner (Du et al., 2021; Gao, 2022). This feature selection method is intuitive in nature and does not well solve the problem of correlation existing between features. Second, feature selection methods such as PCA, XGboost, and LASSO are used to construct the models (Li et al., 2014; Dong dt al., 2018; Kou et al., 2021). Some of the methods in this class of models can retain the features related to the target outcome better, but cannot remove the redundant features. Some methods are able to screen out redundant features, but do not consider the accuracy of prediction of the target outcome. Thus, it is undoubtedly of great theoretical and practical value to construct an effective model and conduct empirical analysis for the dual characteristics of FDP data.

In summary, there are still issues that need to be further explored in FDP research. The occurrence of financial crises is not only caused by internal factors of the company, but also requires consideration of economic, political and social factors. The current selection of indicators mostly stays at the stage of single-source data, and there is no comprehensive and integrated indicator system established, and most of these indicators are numerical in nature. Second, although some of the studies extend the data sources based on machine learning theory and consider text-based variable features, there are problems of dimensional disaster of features and interactions between features in multiple data sources, which require reasonable and accurate feature screening. To address the above issues, our paper aims to conduct a systematic study on FDP of listed companies based on high-dimensional multi-source data. It includes four processes: data sampling, index system establishment, feature selection and financial status classification, and introduces three information sources: internal company, external market and online public opinion, which can reflect the financial status of listed companies systematically and comprehensively. The feature selection based on MRMR-SVM-RFE method can reduce the impact of dimensional catastrophe on prediction accuracy and obtain the feature subset with the highest correlation with the target classification to construct a multi-source listed company FDP model. The comparison of empirical studies of Chinese A-share manufacturing listed companies shows the effectiveness of the method in improving the accuracy of prediction. It provides more comprehensive information for investors and government regulators, and more financial decision basis for corporate financial managers of listed companies in China.

## 2 Indicator system and method

### 2.1 Variable selection

2.1.1 The construction of indicator system

The construction of a reasonable and perfect indicator system is the prerequisite for accurate analysis of the financial crisis situation of listed companies. The merit of the constructed indicator system determines whether it can accurately reflect the company's financial situation, thus affecting the accuracy of the evaluation of the FDP model. Factors affecting the financial situation of a company can be divided into two main categories: internal factors and external factors (Sri, 2016). Internal factors are from the micro level to analyze the internal situation of each enterprise, mainly reflected in several aspects such as financial indicators, earnings management, and corporate governance. External factors are from the macro level, studying the impact of economic environment, legal policies, industry risks, etc. on the financial status of enterprises. The principles of indicator selection were referred to Alfaro et al (2008). The selection of FDP indicators in this study was based on three principles: the selected indicators had been validated in previous studies; the selected indicators were available; and the selected indicators were able to meet the needs of this study. This paper establishes a multi-source data indicator system based on traditional financial indicators, taking internal and external factors into consideration, and the structure diagram is shown in Fig.1.

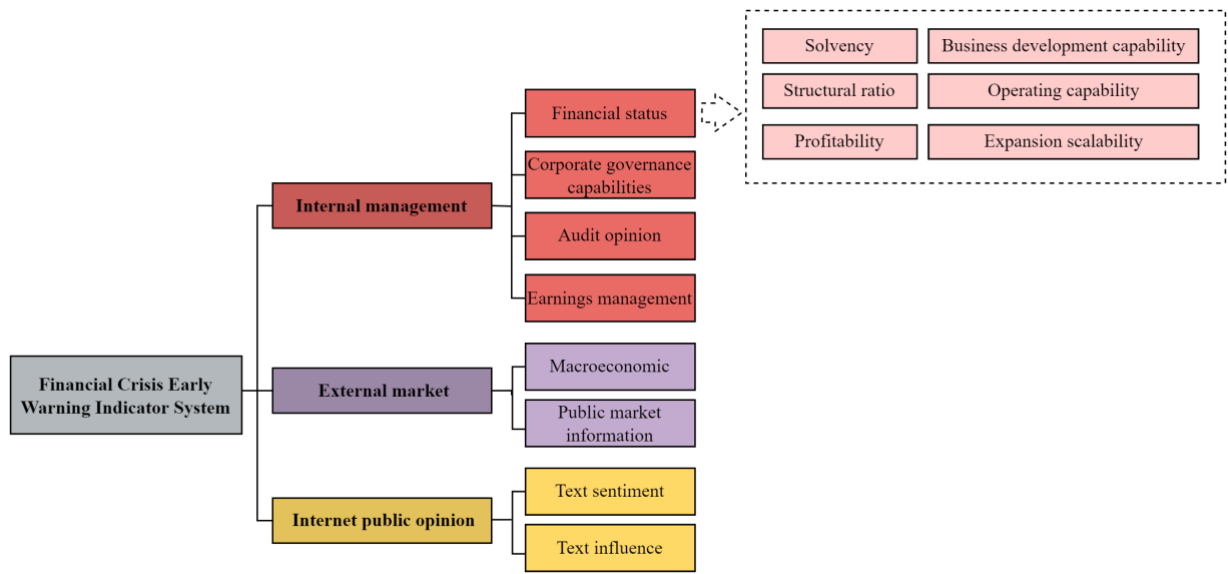

**FIGURE 1** Multi-source data FDP indicator system

(1) Internal management indicators.

The internal management indicators are based on the traditional financial reflecting the financial situation, and introduce earnings management indicators, corporate governance capacity indicators and audit opinions. As the basic framework for predicting the company's financial crisis, traditional financial indicators can provide a good indication of a company's financial health in terms of its financial position and operating results. We considered the comprehensiveness, scientificity, and fairness of indicator selection (Ma et al., 2019), and selected a total of 28 traditional financial indicators from six parts: solvency, business development capacity, structural ratio, profitability, operating capacity, and business expansion capacity. With the increasing improvement of the FDP indicator system, it has been found that the level of corporate governance is positively correlated with corporate performance and significantly negatively correlated with the probability of corporate default (Ali et al., 2018). When the size of the company's management increases, there are checks and balances and mutual supervision of business decisions between the managers of the company, which makes the company's decisions more complete, the financial position and operating results more solid, and the less likely to fall into financial crisis.

In addition, audit opinion and corporate earnings management can reflect the authenticity of corporate financial data to a certain extent and reflect the true financial status of the company from the side. Audit opinion is the opinion of the auditor on whether the object of the assurance meets the assurance criteria after completing the audit (Li, 2019), which reflects the possibility of fraud in the company's financial statements to a certain extent. According to generally accepted accounting principles, the indicator of the added audit opinion can be calculated by Equation (1).

$$Audittyp = \begin{cases} 0, \text{standard unqualified opinion} \\ 1, \text{others} \end{cases} \quad (1)$$

Earnings management indicators need to be further calculated. Currently, companies manage their earnings by manipulating accruals, which are divided into two categories according to whether they are manipulable or not. Non-manipulable accruals are determined by the operating conditions of the company and cannot be managed by the company's management, so earnings management is mainly conducted for manipulable accruals. We draw on the improved Jones model proposed by Dechow et al (Li, 2019) to calculate corporate accruals for earnings management as an indicator of earnings management.

According to the improved Jones model, the company's manipulable accruals earnings management $DAP_t$ in period $t$ is shown in Equation (2):

$$DAP_t = \frac{TA_t}{A_{t-1}} - NDA_t \tag{2}$$

where $A_{t-1}$ represents the ending amount of the company's total assets in period $t-1$; $TA_t$ represents the total accrued profit of the company in period $t$; $NDA_t$ represents the non-manipulable accruals earnings management of the company in period $t$.

The value of $TA_t$ in Equation (2) can be calculated by Equation (3):

$$TA_t = NI_t - OCF_t \tag{3}$$

where $NI_t$ represents the company's operating profit in the $t$ period; $OCF_t$ represents the company's net cash flow from operating activities in the $t$ period. For the calculation method of non-manipulable accrual earnings management $NDA_t$, first of all, the regression is carried out by Equation (4):

$$\frac{TA_t}{A_{t-1}} = \alpha_1 \frac{1}{A_{t-1}} + \alpha_2 \frac{\Delta REV_t}{A_{t-1}} + \alpha_3 \frac{PPE_t}{A_{t-1}} + \mu_t \tag{4}$$

the estimated values $\hat{\alpha}_1$, $\hat{\alpha}_2$, and $\hat{\alpha}_3$ of $\alpha_1$, $\alpha_2$, and $\alpha_3$ are obtained by regression, and then the estimated values are brought into Equation (5) to calculate the non-manipulable accrual earnings management:

$$NDA_t = \hat{\alpha}_1 \frac{1}{A_{t-1}} + \hat{\alpha}_2 \frac{\Delta REV_t - \Delta AR_t}{A_{t-1}} + \hat{\alpha}_3 \frac{PPE_t}{A_{t-1}} \tag{5}$$

In the above process, $\Delta REV_t$ represents the difference between the company's operating income in period $t$ and period $t-1$; $PPE_t$ represents the closing balance of the company's fixed assets in period $t$; $\Delta AR_t$ represents the increase in accounts receivable of the company in period $t$.

The positive and negative of accrual earnings management ($DAP_t$) indicates the direction of management's control of earnings indicators through earnings management, and the absolute value indicates the degree to which earnings information is manipulated and the authenticity of the company's earnings information.

In summary, the 31 variables shown in Table 1 are finally used as alternative internal management indicators.

**Table 1** Internal management indicators

| Categories | | Indicators | |
|---|---|---|---|
| Financial Indicators | Solvency | Current ratio (F1) | Operating cash flow/current liabilities(F4) |
| | | Quick ratio (F2) | Gearing ratio(F5) |
| | | Cash ratio (F3) | Interest coverage multiple(F6) |
| | Business development capability | Total assets growth rate (F7) | Operating income growth rate (F9) |
| | | Net profit growth rate(F8) | — |
| | Structural ratio | Current asset ratio (F10) | Shareholders' equity / fixed assets (F12) |
| | | Fixed asset ratio (F11) | Current Liability Ratio (F13) |
| | Profitability | Net profit margin of total assets (F14) | Operating cost ratio (F17) |
| | | Net profit margin of current assets (F15) | Operating profit margin (F18) |
| | | Net profit rate of fixed assets (F16) | Return on net assets (F19) |
| | Operating capability | Turnover ratio of accounts payable (F20) | Turnover ratio of fixed assets (F23) |

|  |  | Turnover ratio of receivables (F21) | Turnover ratio of total assets (F24) |
|  |  | Turnover ratio of inventories (F22) | — |
|  | Expansion Scalability | Earnings per share (F25) | Capital surplus per share (F27) |
|  |  | Net assets per share (F26) | Net cash flow per share (F28) |
| Corporate governance capabilities | | Number of independent directors/Number of board of directors (GVN) | |
| Audit opinion | | Audit opinion (Audittyp) | |
| Earnings management | | Accrual earnings management (DAP) | |

(2) External Market Indicators

The financial situation of an enterprise is not only related to the internal management of the enterprise. When the social and economic environment changes, the possibility of financial crisis of the enterprise will also change accordingly. When the economic and social environment is favorable, the possibility of financial crisis will also be reduced. The external market indicators selected in this paper are shown in Table 2, which can better reflect the social purchasing power, the social economy and the activity of the public market, changes in labor costs, and changes in the international market.

**Table 2** External market indicators

| Categories | Indicators | |
|---|---|---|
| Macroeconomic indicators | Growth rate of total retail sales of social consumer goods (E1) | RPI growth rate (E6) |
|  | GDP growth rate (E2) | Unemployment rate (E7) |
|  | M1 growth rate (E3) | Growth rate of total import and export (E8) |
|  | M2 growth rate (E4) | Exchange rate (E9) |
|  | CPI growth rate (E5) | — |
| public market information | Total Market Capitalization (M1) | A-shares/total share capital (M2) |

(3) Public opinion indicators

The surge in the number of social network users makes the content of public opinion information have the characteristics of fast dissemination, wide coverage and high degree of freedom (Zhao et

al.,2017). In the new media era, shareholders, as direct beneficiaries of the company's operating conditions, can express their attitudes and opinions on the company's operating behavior and related events through the Internet. Due to the large number of participants and the difficulty of corporate control, their emotional expressions are more authentic compared to internal company information. Public opinion indicators can be reflected by two aspects: text sentiment and text influence. Text sentiment is the sentiment tendency and specific score of the text obtained by analyzing the relevant public opinion text information. The text influence can be calculated through reading, likes, comments and other indicators that can reflect the validity and identity of the text, and the final public opinion index value is obtained by weighting the text influence.

To calculate the sentiment value of the text, first use the crawler technology to collect the online comment information of the shareholders of the sample company to form the online public opinion text, including the content of the post, the reading volume and the comments volume. With the help of jieba word segmentation and stop word removal, the method based on sentiment dictionary is used to convert the text data of posts into numerical data reflecting sentiment tendency, and the sentiment score of each text data is obtained. The text influence is calculated by the reading volume and the comment volume, and the text sentiment score is weighted and summed, and the total score obtained is used as the evaluation standard of the company's network public opinion indicator. The sentiment dictionary in the above processing process selects the existing NTUSD dictionary and CFSD Chinese financial dictionary.

The calculation formula of text influence is shown in Equation (6):

$$emotion = \sqrt[4]{Looks} + \sqrt[2]{Comment} \qquad (6)$$

where $Looks$ and $Comment$ represent the number of reads and comments of the post.

2.1.2 Output variables

There is still considerable disagreement internationally about the criteria for defining a financial crisis, and there is no clear cut-off line to indicate whether a company is in financial crisis or in financial health. The definition of financial crisis in China mainly focuses on listed enterprises, and it is believed that whether an enterprise is in financial crisis depends on whether the listed enterprise is "Special Treatment" ("ST") by the Securities and Futures Commission. In 1988, the Shenzhen and Shanghai Stock Exchanges decided to introduce a "special treatment" policy for listed companies, whereby companies with financial or other abnormalities would be given special treatment, and when a company is ST because of an abnormal financial situation, it is considered to be in financial crisis.

**2.2 Feature selection**

The multi-source data FDP model refers to the use of multiple data sources simultaneously, through data fusion and thus improving the accuracy of prediction. However, more data sources are not better; multiple data sources make the risk of dimensional disasters increase, and there may be interactions between variables from different data sources, which need to be reduced by appropriate feature selection (Luo & Wang, 2020). In this paper, the feature selection method based on MRMR-SVM-RFE is used for feature selection to obtain the feature subset with the minimum redundancy and the maximum relevance to the target classification.

2.2.1 MRMR algorithm

MRMR (Max-Relevance and Min-Redundancy) is a feature selection method proposed by Peng et al. (2005) that can use mutual information, correlation, or distance/similarity scores for feature selection. Its aim is to maximize the correlation between features and categorical variables and minimize the correlation between features and features. The feature selection algorithm of MRMR criterion is a Filter method that operates efficiently. In this paper, the MRMR criterion based on mutual information is used to find a maximum correlation and minimum redundancy feature set.

Let $D=\{x_i\}_{n\times K}$ represent the indicator data matrix, where $x_i$ represents the $i$-th indicator feature and $K$ represents the number of sample companies. Let $x_{\cdot K}=\{x_{1,K},\ldots,x_{n,K}\}$ denote all indicator features of the $k$-th sample firm, $x_{i\cdot}=\{x_{i,1},\ldots,x_{i,K}\}$ denote the cross-sample data of the $i$-th indicator feature, and $G=\{1,2,\ldots,n\}$ denote the set of indicator features.

In this paper, we classify the financial status of companies into two categories: financially healthy (non-ST) companies and financially crisis (ST) companies, i.e., the target categorical variable $y\in\{+1,-1\}$, which takes the value of 1 for ST companies and -1 for non-ST companies. The purpose of feature selection is to select a smaller subset of features $S=\{x_i, i=1,2,\ldots,m\}, m<n$. The correlation between the feature subset and the target variable is defined as shown in Equation (7):

$$R_S = \frac{1}{|S|}\sum_{l}\sum_{i\in S} I(l,i) \qquad (7)$$

where $I(l,i)=\sum_{x_{i\cdot}} p(l,x_{i\cdot})\log\frac{p(l,x_{i\cdot})}{p(l)p(x_{i\cdot})}$ is the mutual information between the target variable and feature $i$, which represents the amount of information contained in one random variable about another random variable. The redundancy of feature $i$ in subset $S$ with other features is given by Equation (8):

$$Q_{S,i} = \frac{1}{|S|^2}\sum_{i'\in S, i'\neq i} I(i,i') \qquad (8)$$

In the MRMR algorithm, the redundancy and relevance of features can be combined in various ways. The MRMR feature selection scoring criterion used in our paper is shown in Equation (9):

$$i^* = \arg\max_{i \in S} \frac{R_S}{Q_{S,i}} \qquad (9)$$

2.2.2 SVM-RFE algorithm

The SVM-RFE algorithm (Support Vector Machine Recursive Feature Elimination) was proposed by Guyon et al. (2001) to rank gene features in gene expression data for cancer classification. It is now widely used in gene selection and is a typical Wrapper method. SVM-RFE removes the least important features for the target categorical variable by means of backward recursive elimination, and gives the weight vector $w$ of the ranking score as shown in Equation (10):

$$w = \sum_{k} \alpha_k y_k x_k \qquad (10)$$

where $y_k$ denotes the target classification label of sample $x_k$ and $\alpha_k$ is the Lagrangian multiplier.

2.2.3 MRMR-SVM-RFE algorithm

When the MRMR algorithm is used alone, the execution speed is fast, and the redundant features and irrelevant features can be found quickly. However, the algorithm is a local optimum rather than a global optimum, and the correlation between the filtered features and the target category cannot be determined, and the best accuracy may not be obtained. On the other hand, SVM-RFE has higher accuracy, but has high computational complexity and does not consider redundancy between features.

In this paper, we propose to combine the two algorithms based on their complementary nature to improve the feature selection of SVM-RFE by minimizing the redundancy between relevant features through the MRMR algorithm.

$$r_i = \beta|w_i| + (1-\beta)\frac{R_{S,i}}{Q_{S,i}} \tag{11}$$

In the MRMR-SVM-RFE method, each indicator feature is ranked by a convex combination of SVM-RFE weight vector values and MRMR criteria. For the $i$-th feature, the ranking score $r_i$ is given by Equation (11).

## 3. Empirical Study

### 3.1 Overall process of model building

The overall process of the multi-source data FDP model is shown in Figure 2. The process is divided into four main stages: sample selection and multi-source data collection, data preprocessing, feature selection, model application and evaluation.

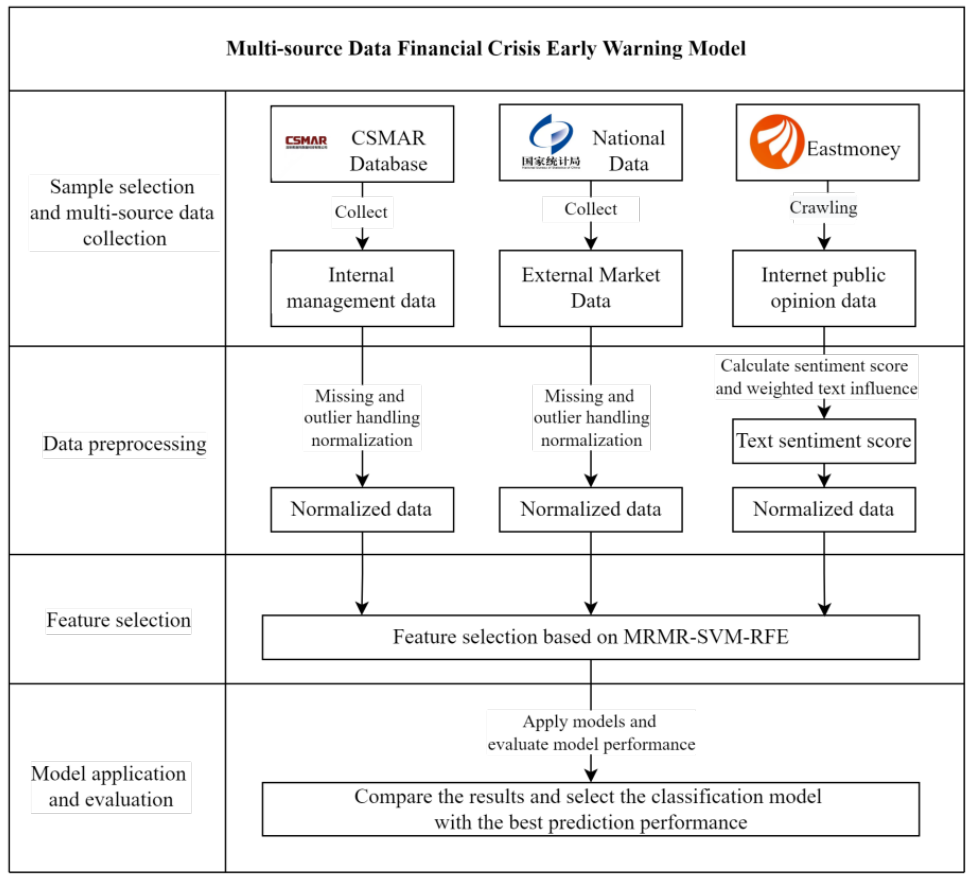

**FIGURE 2** Overall flow of multi-source data FDP model

## 3.2 Sample selection and multi-source data collection

In order to verify the reliability of the model, we select the A-share manufacturing companies listed in China from 2015-2019 as the initial source of the sample for validation. The number of ST-listed companies in Shanghai and Shenzhen is much smaller than the number of ST-listed companies. If we follow the principle of 1:1 matching between financial crisis and financial health, it deviates too much from the actual situation and will lead to an inflated prediction result; if we follow the actual matching principle, it will destroy the balance of the sample. Taking into account the sample balance and practical significance, we selected a total of 75 ST and non-ST companies according to the matching principle of 1:2 ratio of financial crisis to financial health. In the sample of ST companies, we selected 25 companies that were specially treated (ST) for the first time within five years and the reason for special treatment was that the company had lost money for two consecutive years, based on the matching principle of companies in the same industry with similar asset size (Liu, 2019), and for the sample of non-ST companies, 50 companies were randomly selected among companies that had been listed for more than five years and had never been specially treated according to the ratio.

Considering that the company in financial crisis has been in loss for two consecutive years and thus ST, this paper collects index data for the three years before the special treatment. The data of internal management indicators, external market indicators and public opinion indicators are obtained from CSMAR, "National Data" website and Eastmoney.

### 3.3 Raw data pre-processing

The multi-source data FDP model requires data pre-processing for the characteristics of different data sources in order to perform the fusion of different data. For this reason, this section describes the pre-processing process for the various types of data sources involved. In general, in the construction of the multi-source data model, the data pre-processing mainly includes missing value processing, normalization of numerical data and numerical processing of text data due to the increase of data sources and the complexity of data types. In the screening of missing values, the missing values of two indicators, interest coverage multiple (F6) and net profit growth rate (F8), are greater than 30% and are excluded.

For numerical data, since the data values of various characteristics vary greatly, w needs to be normalized in order to eliminate the difference of the range of values on the degree of influence.

For text-based data, since they come from online public opinion and have obvious emotional tendency, their sentiment values need to be calculated. The data after numerical sentiment text also need to be normalized. There are three common normalization methods: the minimum maximum method, the logarithmic method and the statistical method. In this paper, the minimum maximum method is used, and the calculation formula is as in Equation (12).

$$x_i' = \frac{(x_i - X_{i\_min})}{(X_{i\_max} - X_{i\_min})} \tag{12}$$

Among them, $X_{i\_min}$ and $X_{i\_max}$ are the minimum and maximum values of the $i-\text{th}$ feature of the original data set, and $x_i'$ is the normalized data. After normalization, all the original data intervals are $[0,1]$.

### 3.4 Feature selection

Feature selection is a key step in the preprocessing stage of data mining. In high-dimensional classification problems, it is usually used to select a subset of features that are highly related to the target (Du et al., 2019). In our paper, the MRMR-SVM-RFE algorithm is used to select the features of the FDP. For the selection of parameters in Equation (10), we refer to the parameter selection method of Mundra et al. (2010), and select the value of $\beta$ when the accuracy rate is the best from the set $[0, 0.2, 0.4, 0.5, 0.6, 0.8, 1]$. According to the score ranking of variables, the top 20 indicators are selected as the input indicators of the FDP model.

### 3.5 Algorithm selection and result evaluation

To explore the effectiveness of multi-source data metric system and to perform MRMR-SVM-RFE feature selection, LR(Logistic Regression), DT(Decision Tree), BP(Back Propagation),

SVM(Support Vector Machines), RF( Random Forest), GBDT (Gradient Boosting Decision Tree), XGboost (Extreme Gradient Boosting), and seven single classification models and integrated classification models were selected. Using 70% of the samples in the T-3, T-2 and T-1 datasets as the training set and 30% of the samples as the test set, the feature set obtained through MRMR-SVM-RFE feature selection and the unprocessed feature set were used respectively. The original set of features for feature selection to evaluate the classification results of multiple models. Our paper conduct 10 experiments respectively and calculate the average result as the final result to avoid overfitting and improve the robustness of the result.

In order to accurately evaluate the performance of the built model, our paper selects a total of 4 indicators as evaluation criteria, which are described in detail as follows: After the prediction of the data in the dataset, a confusion matrix can be obtained, as shown in Table 3.

**Table 3** Confusion matrix of results

| Predicted value | Real Value | |
|---|---|---|
| | 1 | 0 |
| 1 | TP | FN |
| 0 | FP | TN |

TP (True Positive): The number of samples that are actually in financial crisis and are accurately identified;

FP (False Positive): the number of samples that are actually in financial crisis but are predicted to be in financial health.

FN (False Negative): the number of samples that are actually financially healthy but are predicted to be financially healthy.

TN (True Negative): the number of samples that are actually financially healthy and are accurately predicted.

(1) Precision rate (the probability that a financial crisis can be accurately identified)

$$Precision = \frac{TP}{TP+FP} \tag{13}$$

(2) Recall rate (the correct probability of being predicted as a financial crisis)

$$Recall = \frac{TP}{TP+FN} \tag{14}$$

(3) F1 value

$$F1\text{-}score = 2 \times \frac{\frac{TP}{TP+FP} \times \frac{TP}{TP+FN}}{\frac{TP}{TP+FP} + \frac{TP}{TP+FN}} \tag{15}$$

(4) Accuracy rate

$$Accuracy = \frac{TP+TN}{TP+TN+FN+TN} \tag{16}$$

To observe the prediction effect of different parameter values and different classification algorithms, we conducted many experiments, and obtained the accuracy results of the data sets with different algorithms and three-time spans shown in Table 4-6.

**Table 4** Comparison of prediction effects of three data sets with different parameters in T-3 years

| β | Indexes | Models | | | | | | |
|---|---|---|---|---|---|---|---|---|
| | | LR | DT | BP | SVM | RF | XGboost | GBDT |
| 0 | Precision | 0.8250 | 0.8625 | 0.8625 | 0.8000 | 0.7875 | 0.7750 | **0.8875** |
| | Recall | 0.7106 | 0.7278 | 0.7944 | **0.8125** | 0.7778 | 0.7750 | 0.7897 |
| | F1-score | 0.7626 | 0.7889 | 0.8265 | 0.8058 | 0.7824 | 0.7750 | **0.8352** |
| | Accuracy | 0.8217 | 0.8391 | **0.8739** | 0.8652 | 0.8478 | 0.8435 | 0.8783 |
| 0.2 | Precision | **0.9000** | **0.9000** | 0.8875 | 0.8500 | 0.8000 | 0.7875 | **0.9000** |
| | Recall | 0.8008 | 0.7819 | **0.8278** | 0.8228 | 0.8000 | 0.7681 | 0.7911 |
| | F1-score | 0.8470 | 0.8365 | **0.8559** | 0.8343 | 0.8000 | 0.7772 | 0.8418 |
| | Accuracy | 0.8870 | 0.8778 | **0.8957** | 0.8826 | 0.8609 | 0.8435 | 0.8826 |
| 0.4 | Precision | **0.8875** | 0.8500 | 0.8500 | 0.8375 | 0.7875 | 0.7750 | 0.8750 |
| | Recall | 0.7567 | 0.7406 | **0.8236** | 0.8181 | 0.7885 | 0.7569 | 0.7778 |
| | F1-score | 0.8163 | 0.7907 | **0.8352** | 0.8272 | 0.7874 | 0.7654 | 0.8235 |
| | Accuracy | 0.8609 | 0.8435 | **0.8826** | 0.8783 | 0.8522 | 0.8348 | 0.8696 |
| 0.6 | Precision | **0.8750** | 0.8375 | 0.8375 | 0.8250 | 0.7875 | 0.7625 | 0.8625 |

|   | Recall | 0.7467 | 0.7450 | **0.8111** | 0.7861 | 0.7778 | 0.7542 | 0.7589 |
|   | F1-score | 0.8052 | 0.7881 | **0.8227** | 0.8044 | 0.7824 | 0.7581 | 0.8072 |
|   | Accuracy | 0.8522 | 0.8435 | **0.8739** | 0.8609 | 0.8478 | 0.8304 | 0.8565 |
| 0.8 | Precision | **0.8625** | 0.7875 | 0.8375 | 0.8125 | 0.7750 | 0.7583 | 0.8500 |
|   | Recall | 0.7439 | 0.6922 | **0.7986** | 0.7736 | 0.7653 | 0.7238 | 0.7644 |
|   | F1-score | 0.7979 | 0.7366 | **0.8169** | 0.7919 | 0.7699 | 0.7393 | 0.8042 |
|   | Accuracy | 0.8478 | 0.8043 | **0.8696** | 0.8522 | 0.8391 | 0.8161 | 0.8565 |
| 1 | Precision | **0.8500** | 0.7625 | 0.8250 | 0.7875 | 0.7750 | 0.7500 | 0.8375 |
|   | Recall | 0.7328 | 0.6944 | **0.8065** | 0.7583 | 0.7556 | 0.6917 | 0.7444 |
|   | F1-score | 0.7861 | 0.7265 | **0.8146** | 0.7721 | 0.7647 | 0.7191 | 0.7882 |
|   | Accuracy | 0.8391 | 0.8000 | **0.8696** | 0.8391 | 0.8348 | 0.7957 | 0.8435 |

Table 5 Comparison of prediction effects of the three data sets with different parameters in T-2 years

| β | Indexes | Models | | | | | | |
|---|---|---|---|---|---|---|---|---|
|   |   | LR | DT | BP | SVM | RF | XGboost | GBDT |
| 0 | Precision | 0.8875 | 0.9000 | **0.9264** | 0.9139 | 0.9000 | 0.8875 | 0.9125 |
|   | Recall | 0.7567 | 0.7667 | **0.8558** | 0.8447 | 0.8017 | 0.8092 | 0.8225 |
|   | F1-score | 0.8163 | 0.8275 | **0.8873** | 0.8756 | 0.8469 | 0.8455 | 0.8638 |
|   | Accuracy | 0.8609 | 0.8696 | **0.9174** | 0.9087 | 0.8870 | 0.8870 | 0.9000 |
| 0.2 | Precision | 0.9000 | 0.9250 | 0.9625 | 0.9500 | 0.9250 | 0.9375 | **0.9750** |
|   | Recall | 0.8008 | 0.8336 | **0.9083** | 0.9069 | 0.8433 | 0.8350 | 0.8506 |
|   | F1-score | 0.8470 | 0.8756 | **0.9330** | 0.9264 | 0.8807 | 0.8822 | 0.9070 |
|   | Accuracy | 0.8870 | 0.9087 | **0.9522** | 0.9478 | 0.9130 | 0.9130 | 0.9304 |
| 0.4 | Precision | 0.8875 | 0.9125 | **0.9500** | 0.9375 | 0.9125 | 0.9000 | 0.9375 |
|   | Recall | 0.7567 | 0.8208 | **0.9069** | 0.8944 | 0.8544 | 0.8531 | 0.8581 |
|   | F1-score | 0.8163 | 0.8640 | **0.9264** | 0.9139 | 0.8799 | 0.8733 | 0.8931 |
|   | Accuracy | 0.8609 | 0.9000 | **0.9478** | 0.9391 | 0.9130 | 0.9087 | 0.9217 |
| 0.6 | Precision | 0.8750 | 0.9000 | **0.9250** | 0.9000 | 0.8875 | 0.8875 | 0.9125 |
|   | Recall | 0.7467 | 0.8097 | **0.8931** | 0.8903 | 0.8481 | 0.8383 | 0.8508 |
|   | F1-score | 0.8052 | 0.8522 | **0.9073** | 0.8941 | 0.8661 | 0.8609 | 0.8793 |
|   | Accuracy | 0.8522 | 0.8913 | **0.9348** | 0.9261 | 0.9043 | 0.9000 | 0.9130 |
| 0.8 | Precision | 0.8625 | 0.8875 | **0.9125** | 0.8875 | 0.8750 | 0.8875 | **0.9125** |
|   | Recall | 0.7439 | 0.8083 | **0.8792** | 0.8569 | 0.8365 | 0.8189 | 0.8314 |
|   | F1-score | 0.7979 | 0.8456 | **0.8949** | 0.8713 | 0.8534 | 0.8507 | 0.8690 |
|   | Accuracy | 0.8478 | 0.8870 | **0.9261** | 0.9087 | 0.8957 | 0.8913 | 0.9043 |
| 1 | Precision | 0.8500 | 0.8750 | **0.9125** | 0.8875 | 0.8875 | 0.9000 | **0.9125** |
|   | Recall | 0.7328 | 0.7958 | 0.8233 | 0.8278 | 0.7994 | 0.8114 | **0.8606** |
|   | F1-score | 0.7861 | 0.8331 | 0.8637 | 0.8559 | 0.8404 | 0.8520 | **0.8845** |
|   | Accuracy | 0.8391 | 0.8783 | 0.9000 | 0.8957 | 0.8826 | 0.8913 | **0.9174** |

Table 6 Comparison of prediction effect rates for the three data sets with different parameters in year T-1

| β | Indexes | Models | | | | | | |
|---|---|---|---|---|---|---|---|---|
| | | LR | DT | BP | SVM | RF | XGboost | GBDT |
| 0 | Precision | 0.9125 | 0.9125 | **0.9514** | 0.9389 | 0.9375 | 0.9000 | 0.9250 |
| | Recall | 0.7791 | 0.7767 | **0.8883** | 0.8572 | 0.8053 | 0.8314 | 0.8350 |
| | F1-score | 0.8390 | 0.8386 | **0.9168** | 0.8940 | 0.8652 | 0.8632 | 0.8763 |
| | Accuracy | 0.8783 | 0.8783 | **0.9391** | 0.9217 | 0.9000 | 0.9000 | 0.9087 |
| 0.2 | Precision | 0.9250 | 0.9375 | **0.9750** | **0.9750** | 0.9375 | 0.9625 | **0.9750** |
| | Recall | 0.8328 | 0.8447 | **0.9306** | 0.9083 | 0.8736 | 0.8697 | 0.8617 |
| | F1-score | 0.8757 | 0.8873 | **0.9515** | 0.9397 | 0.9037 | 0.9123 | 0.9129 |
| | Accuracy | 0.9087 | 0.9174 | **0.9652** | 0.9565 | 0.9304 | 0.9348 | 0.9348 |
| 0.4 | Precision | 0.9250 | 0.9250 | **0.9625** | 0.9500 | 0.9375 | 0.9125 | 0.9375 |
| | Recall | 0.7716 | 0.8328 | **0.9194** | 0.8958 | 0.8906 | 0.8656 | 0.8692 |
| | F1-score | 0.8404 | 0.8757 | **0.9389** | 0.9205 | 0.9100 | 0.8858 | 0.8990 |
| | Accuracy | 0.8783 | 0.9087 | **0.9565** | 0.9435 | 0.9348 | 0.9174 | 0.9261 |
| 0.6 | Precision | 0.9000 | 0.9125 | **0.9375** | 0.9125 | 0.9000 | 0.8875 | 0.9250 |
| | Recall | 0.7589 | 0.8208 | **0.9056** | 0.8903 | 0.8842 | 0.8472 | 0.8522 |
| | F1-score | 0.8229 | 0.8640 | **0.9198** | 0.9007 | 0.8895 | 0.8662 | 0.8859 |
| | Accuracy | 0.8652 | 0.9000 | **0.9435** | 0.9304 | 0.9215 | 0.9043 | 0.9174 |
| 0.8 | Precision | 0.9000 | 0.9000 | **0.9375** | 0.9000 | 0.9125 | 0.9000 | 0.9125 |
| | Recall | 0.7494 | 0.8194 | **0.8819** | 0.8681 | 0.8517 | 0.8300 | 0.8411 |
| | F1-score | 0.8168 | 0.8574 | **0.9081** | 0.8831 | 0.8792 | 0.8624 | 0.8742 |
| | Accuracy | 0.8605 | 0.8957 | **0.9348** | 0.9174 | 0.9130 | 0.9000 | 0.9087 |
| 1 | Precision | 0.8750 | 0.8875 | **0.9375** | 0.9000 | 0.8875 | 0.9125 | 0.9125 |
| | Recall | 0.7466 | 0.8069 | **0.8633** | 0.8403 | 0.8092 | 0.8128 | 0.8322 |
| | F1-score | 0.8043 | 0.8449 | **0.8977** | 0.8684 | 0.8455 | 0.8587 | 0.8690 |
| | Accuracy | 0.8522 | 0.8870 | **0.9261** | 0.9043 | 0.8870 | 0.8957 | 0.9043 |

Overall, in all three datasets, the prediction effect is better when the parameter value $\beta$ is 0.2 than other parameter values. When the parameter $\beta$ takes the value of 0.2, the best prediction effect is achieved on the T-1 data set using the BP classification model, and the prediction accuracy reaches 0.9652, and the precision rate is the highest value of 0.9750 in the experimental results, which can identify the listed companies in financial crisis more accurately. The worst prediction effect was observed when the parameter value was 1 (feature selection using only the SVM-RFE algorithm alone) and when the parameter value was 0 (feature selection using only the MRMR algorithm alone).

Specifically, on the T-3 dataset, it can be seen from Table 4 that the highest alert accuracy occurs in the BP classification algorithm (0.8739) and the highest recall occurs in the SVM model (0.8125) when $\beta$ takes the value of 0. The GBDT algorithm performs better on the two evaluation measures of accuracy (0.8875) and F1 value (0.8352). When the value of $\beta$ is 0.2, the prediction effect is the best, and the highest recall rate, F1 value and accuracy rate all appear in the BP classification algorithm, which are 0.8278, 0.8559, and 0.8957, respectively. The highest precision rate appears in the three classification algorithms LR, DT, and GBDT, which is 0.9000, an improvement of 0.0125 compared to the parameter of 0. The peak values of prediction accuracy, recall, F1 value and precision are found when $\beta$ is taken as 0.2, and the prediction effect decreases when 0.4, 0.6, 0.8 and 1 are taken in order. BP algorithm performs well in recall, F1 value and accuracy, and LR and GBDT algorithms perform better in terms of precision. The prediction effect is worse when $\beta$ is 1 than when the value is 0.

On the T-2 dataset, it can be seen from Table 5 that the two top performing classification algorithms are the BP classification algorithm and the GBDT classification algorithm. The T-2 dataset has improved in all four evaluation measures compared to the T-3 dataset, and all evaluation metrics reach their highest values when $\beta$ takes the value of 0.2, which are 0.9750, 0.9083, 0.9330, and 0.9522, where the highest precision rate is derived from the performance of the GBDT classification algorithm, and the highest recall rate, highest F1 value, and highest accuracy rate are all based on the performance of the BP classification algorithm's performance.

On the T-3 data set, it can be seen from Table 6 that the best-performing classification algorithm is still the BP algorithm, and the result is the best when the value of $\beta$ is 0.2. The performance of the BP classification algorithm on the T-1 dataset is improved compared to the T-2 dataset. When the value of $\beta$ is 0.2, the precision, recall, F1 value and accuracy of the BP classification model on the T-1 dataset are improved by 0.0125, 0.0223, 0.0185, and 0.013 compared with the T-2 dataset.

All in all, when using the MRMR-SVM-RFE algorithm for feature selection, attention should be paid to the influence of parameter value selection on the prediction accuracy. Selecting appropriate parameter values can improve the accuracy and other evaluation measures by more than 0.2 at most. In this experiment, the prediction effect is optimal when the $\beta$ value is 0.2. Secondly, the prediction effects of different classification models are quite different. The BP classification model showed a large prediction advantage in the three datasets. Meanwhile, datasets with different time spans show different prediction effects. Compared with the T-2 and T-3 datasets, the T-1 dataset has better prediction performance, and the T-3 year has the lowest prediction accuracy. The prediction accuracy rates of different time periods are significantly different. The closer to the time when the listed company is specially treated, the easier it is to judge the possibility of the listed company being specially treated. On the basis of this conclusion, the feature selection results when the $\beta$ value is 0.2 (Table 7) and the comparison results of different index systems of the BP algorithm on the T-1 data set are listed (Figure 3).

**Table 7** Feature selection results

| Categories | | T-3 | T-2 | T-1 |
|---|---|---|---|---|
| Internal indicators | Solvency | F4、F5 | F4、F5 | F1、F3 |
| | Business Development capability | F7、F9 | F7 | F7、F9 |
| | Structural ratio | F10、F12、F13 | F12、F13 | F10、F12、F13 |
| | Profitability | F17、F19 | F14、F15、F17、F18 | F15、F17 |
| | Operating capability | F20、F21、F23 | F20、F21、F24 | F21、F22 |
| | Expansion scalability | F27、F28 | F25 | F25、F28 |

|  | Corporate governance capabilities | GVN | GVN | GVN |
|---|---|---|---|---|
|  | Audit opinion | Audittyp | Audittyp | Audittyp |
|  | Earnings management | DAP | DAP | DAP |
| External Indicators | Macroeconomic | —— | E3、E5、E8 | E3、E5、E9 |
|  | Public market information | M1、M2 | —— | —— |
| Internet public opinion |  | emotion | emotion | emotion |

Note: "——" means no variables were selected under this category

Table 7 gives the results of the top 20 features with scores selected from the three-year dataset of T-3, T-2, and T-1 when $\beta=0.2$. Among the selected internal indicator features, financial indicators still account for the majority of the selected features, covering six aspects: solvency, business development capability, structural ratio, profitability, operating capability, and expansion capability. Among the indicators reflecting solvency, operating cash flow/current liabilities (F4) and gearing ratio (F5) are selected for both the T-3 and T-2 data sets, while current ratio (F1) and cash ratio (F3) are selected for the T-1 year. The cash flow from operating activities to current liabilities ratio (F4) reflects whether debts can be repaid in a timely manner, and both the current ratio (F1) and cash ratio (F3) reflect the liquidity of the enterprise. It indicates that the liquidity of the enterprise can directly induce the financial crisis, and the enterprise should focus on its own financial liquidity. The selection of business development capability is nearly the same for the three datasets. The growth rate of total assets (F7) is selected for all three datasets, and the growth rate of operating income (F9) is additionally selected for the T-3 and T-1 datasets, which can better reflect the financial status of enterprises. Among the indicators reflecting structural ratios, the indicator of fixed assets ratio (F11) was not selected in any of the three data sets, so the idle capital status of the company is not effective in determining the financial crisis of the enterprise. Among the indicators reflecting profitability, the operating cost ratio (F17) reflects the company's ability to control costs and can be used as an effective indicator for FDP. Among the operating capability indicators, accounts receivable turnover (F21) better reflects the company's financial situation. The easier it is to collect accounts receivable, the less likely it is that the company will have a financial crisis. Companies should also pay more attention to the inventory turnover rate (F22). The faster the inventory turnover rate is, the more liquid the company's capital is, which is related to the company's solvency. The two indicators of enterprise expansion capability, earnings per share (F25) and net cash flow per share (F28), both reflect the company's financial situation better. Among the internal indicators other than traditional financial indicators, corporate governance capability (GVN), audit opinion (Audittyp), and earnings management indicator (DAP) are all selected simultaneously by T-3, T-2, and T-1 datasets, and the inclusion of these three indicators can improve the accuracy.

For the external indicators, only the characteristics from the public market information are selected for the year T-3, indicating that the depressed development of the stock market is likely to be the source of the crisis in the company's financial situation, while factors such as the country's monetary policy, the population's consumption and the import/export trade situation are important reasons for the company's continuous losses in the years T-2 and T-1 and thus for the special treatment. The online public opinion indicators in the three datasets were selected simultaneously, indicating that the inclusion of online public opinion indicators can improve the accuracy. A total of nine variables including online public opinion indicators are repeatedly selected in Table 7 by the three time-span datasets, and these feature variables can be considered as the most capable variables for FDP, covering traditional financial indicators, governance capability, audit opinion, surplus management, and online public opinion.

Next, we conducted a comparison experiment on the accuracy of different financial indicator systems, and the feature selection was still performed at the parameter $\beta=0.2$, based on the BP classification algorithm. We compare and analyze the prediction effects in three scenarios: the traditional financial indicator system, the original multi-source data indicator system fusing three data sources, and the multi-source data indicator system after MRMR-SVM-RFE feature selection. The comparison results are shown in Figure 3.

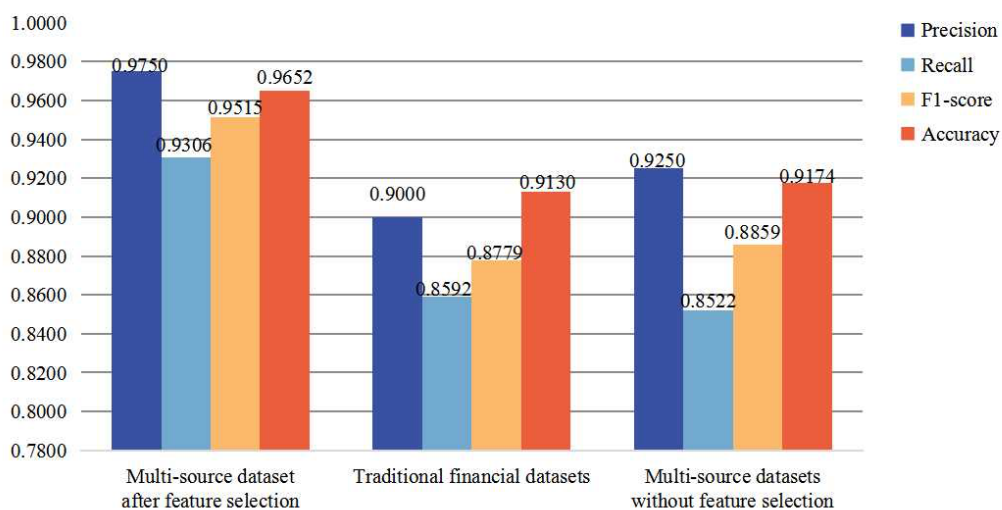

**FIGURE 3** Comparison of prediction effects of different indicator systems based on BP classification algorithm at β=0.2 on T-1 data set

In addition, the comparison results in Figure 3 show that compared with the original multi-source data indicator system using the traditional financial indicator system and fusing three data sources to build the model, the multi-source data indicator system after MRMR-SVM-RFE feature selection can improve the prediction effect more obviously and perform well in each evaluation index. Compared with the multi-source dataset without feature selection, the precision, recall, F1 value and accuracy of the multi-source dataset after feature selection improved by 0.0750, 0.0714, 0.0736 and 0.0522. The difference between the model prediction effect of the traditional financial indicator system and the original multi-source data indicator system that fuses three data sources is not significant. This is because simply fusing three data sources together will cause data redundancy, and some indicators that are not related to the financial crisis are retained in the model without being eliminated, thus leading to a lower prediction effect. In addition, the multicollinearity among the factors also affects the accuracy.

## 4 Conclusion

Our study combines the experience of previous studies and the basic national conditions of China, and establishes a multi-source data listed company FDP indicator system for the situation of Chinese listed companies, covering a total of 43 indicator features from three sources: internal management, external market and online public opinion. To solve the redundancy problem of multi-source indicator features, we used MRMR-SVM-RFE feature selection algorithm to extract valuable features after data pre-processing. Finally, we evaluate the results using seven single classification models and integrated classification models, using the selected features as input factors for the classification models. The results confirm that the FDP model based on the MRMR-SVM-RFE feature selection algorithm has better classification performance and that companies can use the BP classification model for FDP, which is applicable to listed companies for FDP. In response to the conclusions obtained from the empirical analysis, the following recommendations are made to managers of listed companies:

First, the occurrence of a financial crisis in an enterprise is precursory and predictable. Enterprise management can prevent risks by establishing a FDP system. But first of all, financial managers should ensure the timeliness and authenticity of financial information, pay attention to every detail of deviation, and ensure that the FDP system is efficient and accurate, which can better help the enterprise to discover the bad financial situation. The enterprise management should adjust the corporate governance structure at the right time to ensure mutual coordination, mutual supervision and checks and balances among the managers, establish a standardized corporate governance structure and form a mechanism of checks and balances of rights.

Second, enterprise management should pay attention to analyzing the pros and cons of financial information manipulation behaviors such as earnings management and use them carefully. Although surplus management can briefly ensure corporate interests, the distortion of financial information caused by excessive surplus management behavior will eventually lead to the outbreak of a financial crisis. When an enterprise's earnings management behavior is abnormal, further attention should be paid to whether the enterprise is facing higher financial risks and promptly make adjustments. Regulators should also pay attention to corporate manipulation of financial information, strengthen market supervision, and improve the penalty mechanism to ensure the quality of financial information as much as possible.

Thirdly, the management of the enterprise should pay full attention to the changes of internal and external environment. The management cannot focus only on the internal conditions of the enterprise and cannot conclude that all indicators are normal because of the excellent conditions of one or two of them, but should make an objective and comprehensive analysis of the internal environment of the enterprise. Enterprises in the big data environment should actively use big data technology to monitor the external market environment and network public opinion, guide the direction of public opinion development in real time, and be alert to the impact of changes in the external market environment and network public opinion on the company's operation.

There are still some limitations in the research of this paper. Although this paper more comprehensively integrates data from three data sources: internal management, external market and network public opinion, there are many other factors that affect the financial status of enterprises, such as disclosure of information in financial reports of listed companies and human resource management indicators. Incorporating more relevant indicators into the multi-source data financial indicator system may further improve the accuracy of prediction. The second is the parameter selection problem in conducting MRMR-SVM-RFE feature selection. Our paper is to compare the accuracy by assigning intervals to the parameters and selecting the parameter value with the highest accuracy, and the selected parameter value is not certainly the optimal parameter value. Therefore, the future work of this paper will further optimize the FDP index system and parameter value selection algorithm.